\documentclass[11pt]{article}
\usepackage{epsfig}
\title{\bf Gluon emission in interaction of two reggeons}
\author{M.A.Braun, S.S.Pozdnyakov, M.Yu.Salykin, M.I.Vyazovsky\\
{\it S.Petersburg State University, Russia}}
\begin{document}

\maketitle

\def\beq{\begin{equation}}
\def\eeq{\end{equation}}
\def\disc{{\rm Disc}}
\def\lra{\leftrightarrow}
\def\ep{\epsilon}
\def\ta{\tau^{(1)}}
\def\tb{\tau^{(2)}}
\def\tc{\tau^{(3)}}
\def\aa{a^{(1)}}
\def\ab{a^{(2)}}
\def\qa{q_{1+}}
\def\qb{q_{2+}}
\def\ra{r_{1-}}
\def\rb{r_{2-}}
\def\bt{\bar{t}}

{\bf Abstract}

The vertex is constructed for gluon production in  interaction of two
reggeons coupled to  projectiles and two reggeons coupled to  targets.
The vertex can be used to build cross-sections for collisions of two
pairs of nucleons in AA scattering. Transversality of the constructed
vertex is demonstrated as well as its good behaviour at large
longitudinal momenta necessary for applications. Poles at zero values
of longitudinal momenta are discussed and it is found
that they remain in the amplitudes unlike in the case of a single
projectile.

\section{Introduction}
In the QCD hadronic interactions at high energies in the Regge kinematics,
when the transferred transverse momenta are much smaller than energies,
can be described by the interaction of normal gluons with reggeized
ones ("reggeons"). The latter combine into pomerons coupled to participant
colourless projectiles and targets. The simplest case is the collision
of the highly virtual gluon with a hadron or a nucleus. It has been studied
long ago and solved by formulation of the BFKL ~\cite{BFKL1,BFKL2} and
Balitski-Kovchegov ~\cite{bal,kov} equations for the non-integrated
gluon densities, which allow to calculate
the relevant cross-sections. With some ingenuity this approach can be
generalized to pp or pA scattering.
However, nucleus-nucleus collisions present a more difficult problem.
In the case of heavy nuclei the total cross-sections can be treated within
the effective pomeron interaction formalism ~\cite{bra1}. For the inclusive
gluon production a general formalism was developed
in ~\cite{gelis1,gelis2,gelis3}
in the framework of the Color Glass Condensate (CGC) approach. The inclusive
cross-sections were expressed via averages of the gluon potentials in the field
of the colliding nuclei, developed in rapidity according to the so-called
JIMWLK functional equations. These averages can only be found by numerical
methods. Several attempts to find analytic expressions for the inclusive gluon
production have lead to only approximate ~\cite{dusling} or partial
and inconclusive results ~\cite{kov1,bal1}.

The BFKL approach presents an alternative (in all probability equivalent,
in principle) way to study this problem. It may produce analytic formulas
for the cross-sections and also allow to study the case of light nuclei,
when the leading contributions appear to be of the subleading order in $N_c$
~\cite{dd} and a direct application of the CGC approach does not seem
to be possible. In the BFKL approach the problem requires knowledge
of amplitudes for gluon production in interaction of many reggeons
coupled to the projectile with many reggeons coupled to the target.

\begin{figure}
\hspace*{2 cm}
\epsfig{file=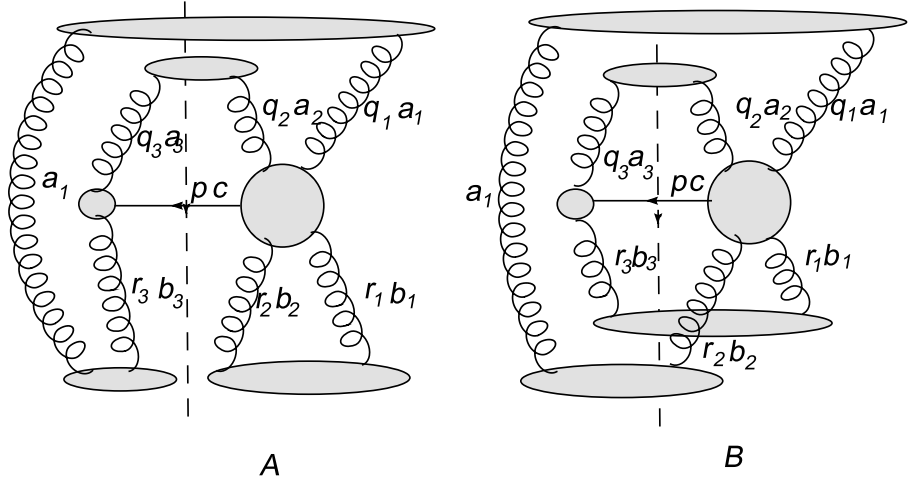, width=12 cm}
\caption{Gluon production by pair of nucleons of the
projectile in collision with a pair of nucleons from the target.
Solid lines correspond to particles, wavy ones to reggeons}
\label{fig1}
\end{figure}

The simplest non-trivial case is production of gluons in interaction
of two reggeons coupled to the projectile with two reggeons coupled to the
target. For heavy nuclei, in the lowest order, this means
interaction of two nucleons from each of the nuclei, as illustrated
in Fig. \ref{fig1}. One observes that apart from the well-known
production amplitude in collision between two colourless objects
there enters a more complicated vertex when the gluon is produced by
a pair of reggeons coupled to projectile and target nucleons
(the RR$\to$RRP vertex where P stands for "particle", that is gluon).

In this paper we study this RR$\to$RRP vertex both with a real
(on-mass-shell) and virtual (off-mass-shell) gluon. The latter is
needed if one wants to construct the vertex for transition of three
reggeons into three reggeons (RRR$\to$RRR), which enters the kernel of the
equation for the odderon or the higher pomeron made of the three
reggeons. Note that for the odderon the RRR$\to$RRR vertex,
integrated over the longitudinal momenta, was derived in ~\cite{odderon}

\begin{figure}
\hspace*{4 cm}
\epsfig{file=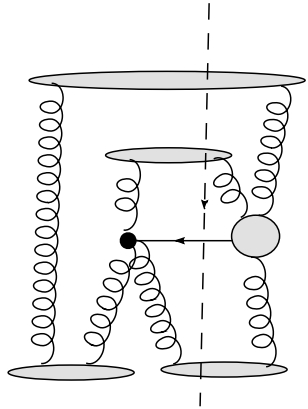, width=4 cm}
\caption{Gluon production by pair of nucleons of the
projectile in collision with a pair of nucleons from the target
from cutting the RR$\to$RRP vertex. The left vertex is induced.}
\label{fig2}
\end{figure}

Note that for collisions of heavy nuclei the contribution shown in
Fig. \ref{fig1} is only a particular term corresponding to interaction
of only two pairs of nucleons. By itself is corresponds only gluon
production in collision of two deuterons. But even in this simple case
one should additionally consider contributions from cutting the vertex
itself, as shown in Fig. \ref{fig2}. As seen from Fig. \ref{fig1},
the immediate application of the
vertex RR$\to$RRP is to give a non-trivial contribution
to the diffractive gluon production in deuteron-proton collisions
(Fig. \ref{fig1},A)

In this paper we do not attempt to calculate this contribution, which
requires a lot more of analytical and numerical effort.
Our aim is to just derive the vertex itself and demonstrate it basic
properties important to is subsequent applications: transversality,
vanishing at large longitudinal momenta and presence or absence of
poles at zero values of the latter.

The main tool for the calculation of the vertex is the Lipatov effective
action ~\cite{lip}, which gives the rules for reggeon-particle
interaction  at a given rapidity and introduces the so-called
induced vertices for this interaction. Some of the full and induced
vertices derived from this action have been already found in
~\cite{bravyaz}.
The induced vertex
RP$\to$RP is new and will be derived here.

The paper is organized as follows. In the next section we derive the
vertex RR$\to$RRP off the mass shell. In Section 3. we demonstrate its
transversality. Section 4 is devoted to the study of the vertex at high
longitudinal momenta. In Section 5 we derive the vertex on the mass-shell.
In Section 6 we investigate the pole singularities at zero values of the
longitudinal momenta. In the last section we make some conclusions.

\section{Vertex RR$\to$RRP with a virtual emitted gluon}
\begin{figure}
\hspace*{2 cm}
\epsfig{file=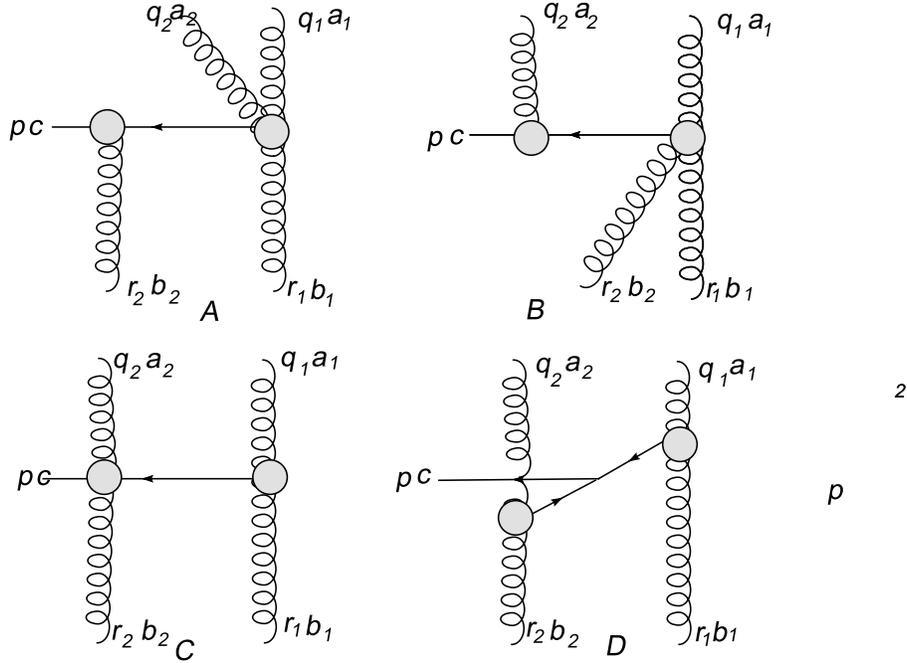, width=12 cm}
\caption{Different contributions to the vertex RR$\to$RRP.
Solid lines correspond to particles, wavy ones to reggeons}
\label{fig3}
\end{figure}

In the framework of the effective action the vertex RR$\to$RRP
is constructed as a sum of four diagrams, shown in Fig.\ref{fig3},~A,...,D
with subsequent symmetrization in the reggeons attached to the projectile
(upper in Fig. \ref{fig3}) and those attached to the target (lower in
Fig. \ref{fig3}). The blobs in Fig. \ref{fig3} denote full (basic plus
induced) vertices. Solid lines denote gluons, wavy ones correspond
to reggeons. We denote momenta and colours of upper reggeons from
right to left as $q_1,a_1$ and $q_2,a_2$ and those for lower reggeons
as $r_1,b_1$ and $r_2,b_2$. The emitted gluon has its momentum,
polarization and colour $p,\mu, c$. The reggeons carry polarization
vectors
$n^{\pm}$ with $n^+_+=n^+_\perp=n^-_-=n^-_\perp=0$, $n^+_-=n^-_+=1$.

\subsection{Fig. \ref{fig3},A}
As is clear from the figure  we already know all the building
blocks for the construction of the vertex. The vertex on the right
RR$\to$RP can be found from the vertex R$\to$RRP, calculated
in ~\cite{bravyaz} after changing the direction
of reggeon propagation and notations of the momenta.
From the same publication one can extract the  vertex on the left
P$\to$RP.

In this way we find the vertex RR$\to$RP on the right in the form

\beq
\bar{V}_\nu=i\frac{f^{b_1a_1e}f^{ea_2d}}{t_1^2+i0}
\Big({\bar A}t_\nu-{\bar B}q_{1\nu}-{\bar C}q_{2\nu}
+{\bar D}n^+_\nu+{\bar E}n^-_\nu\Big),
\label{bv1}
\eeq
where  $t=p+r_2=q_{1}+q_{2}-r_1$, $t_1=q_1-r_1$ and
\[
{\bar A}=3t_-+\frac{r_1^2}{q_{1+}},\ \ {\bar B}=4t_-,\ \
{\bar C}=4t_-+2\frac{r_1^2}{q_{1+}},\]
\[
{\bar D}=-\frac{r_1^2(r_1-q_1)^2}{t_+q_{1+}}
-2t_-\frac{r_1^2}{q_{1+}}-4t_-^2,\]
\beq
{\bar E}=-\Big(-(r_1+q_1)(t+q_2)+q_2^2-q_1^2+(r_1-q_1)^2
+2r_{1-} q_{1+}\Big)+\Big(-2r_{1-}+ \frac{r_1^2}{q_{1+}} \Big)
\Big(t_+ +q_{2+}-\frac{q_2^2}{r_{1-}}\Big).
\label{babcd}
\eeq
Note that this vertex is transversal $(t\bar{V})=0$.

On the left of the diagram Fig. \ref{fig3},A there stands the vertex
R$\to$RP, given by the tensor ~\cite{bravyaz}
\beq
X_{\mu\nu}=-gf^{db_2c}\Big((p+t)_+g_{\mu\nu}+
(p-2t)_\mu n^+_\nu
+(t-2p)_\nu n^+_\mu-n^+_\mu n^+_\nu\frac{r_2^2}{p_+}\Big).
\label{tx}
\eeq
This vertex is not orthogonal to $p$ or $t$ separately.
However, the product $(pXt)=0$. The term $t_{\nu} n^+_{\mu}$ does not
contribute due to the transversality of $\bar{V}_\nu$ and will be dropped.

Multiplying $X$ by $\bar{V}$ on the right we obtain the contribution
to the vertex RR$\to$RRP from the diagram in Fig. \ref{fig3},A:
\beq
{\cal A}_{1\mu}=-g^3C_1\frac{1}{(t_1^2+i0)(t^2+i0)}
\Big(a_\mu{\bar A}-b_\mu{\bar B}-c_\mu{\bar C}+d_\mu{\bar D}+
e_\mu{\bar E}\Big),
\label{atot}
\eeq
where vectors $a,...e$ are
\[a_\mu=p_\mu p_+-n_\mu^+(t^2+p^2),\]
\[b_\mu=2p_+q_{1\mu}+(p-2t)_\mu q_{1+}-n^+_\mu\Big(2(pq_1)+
r_2^2\frac{q_{1+}}{p_+}\Big),\]
\[c_\mu=2p_+q_{2\mu}+(p-2t)_\mu q_{2+}-n^+_\mu\Big(2(pq_2)+
r_2^2\frac{q_{2+}}{p_+}\Big),\]
\[d_\mu=0,\]
\[e_\mu=2p_+n^-_{\mu}+(p-2t)_\mu-n^+_{\mu}
\Big(2p_-+\frac{r_2^2}{p_+}\Big)\]
and the colour coefficient $C_1$ is
\beq
C_1=f^{db_2c}f^{b_1a_1e}f^{ea_2d}.
\eeq

\subsection{Fig. \ref{fig3},B}
In this case the R$\to$RRP vertex on the right can be taken directly
from ~\cite{bravyaz} duly changing the notations. We get for it
\[
V_\nu=i\frac{f^{db_2e}f^{eb_1a_1}}{t_1^2+i0}
\Big\{q_{1+}(4r_1+\bar{t})_\nu-\Big[(q_1+r_1)(\bar{t}-r_2)+r_2^2-r_1^2+
(q_1-r_1)^2+2q_{1+}r_{1-}\Big]n^+_\nu\]\[+
\frac{q_1^2(q_1-r_1)^2}{\bar{t}_-r_{1-}}n^-_\nu+
\Big(2q_{1+}-\frac{q_1^2}{r_{1-}}\Big)
\Big[-2q_{1+}n^-_\nu+(\bar{t}+2r_2)_\nu+
\Big(\bar{t}_- -r_{2-}+\frac{r_2^2}{q_{1+}}\Big)n^+_\nu\Big]\Big\},
\]
where $\bar{t}=q_1-r_1-r_2$.
We rewrite it as
\[
V_\nu=i\frac{f^{db_2e}f^{eb_1a}}{(q_1-r_1)^2+i0}
\Big\{A\bt_\nu+Br_{1\nu}+Cr_{2\nu}+Dn^-_\nu+En^+_\nu\Big\},
\]
where
\[
A=3q_{1+}-\frac{q_1^2}{r_{1-}},\ \ B=4q_{1+},\ \
C=2\Big(2q_{1+}-\frac{q_1^2}{r_{1-}}\Big),\]
\[
D=\frac{q_1^2(q_1-r_1)^2}{\bt_-r_{1-}}-
2q_{1+}\Big(2q_{1+}-\frac{q_1^2}{r_{1-}}\Big),\]
\beq
E=-\Big((q_1+r_1)(\bt-r_2)+r_2^2-r_1^2+
(q_1-r_1)^2+2q_{1+}r_{1-}\Big)+\Big(2q_{1+}-\frac{q_1^2}{r_{1-}}\Big)
\Big(\bar{t}_{-}-r_{2-}+\frac{r_2^2}{q_{1+}}\Big).
\label{abcd}
\eeq

Vertex $\bar{X}$ in the diagram on the left is obtained from
(\ref{tx}) by inversion of reggeons with $p$ and $\bar{t}$ preserved:
\beq
  {\bar X}_{\mu\nu}=
-gf^{da_2c}\Big(2p_-g_{\mu\nu}+
(p-2\bt)_\mu n^-_\nu+ n^-_\mu (\bt-2p)_\nu
-n^-_\mu n^-_\nu\frac{q_2^2}{p_-}\Big).
\label{tbx}
\eeq
Multiplying it by $V$ on the right we get
the contribution
to the vertex RR$\to$RRP from the diagram in Fig. \ref{fig3},B:
\beq
{\cal A}_{2\mu}=-g^3C_2
\frac{1}{(t_1^2+i0)(\bar{t}^2+i0)}
\Big({\bar a}_\mu A+{\bar b}_\mu B+{\bar c}_\mu C+{\bar d}_\mu D+
{\bar e}_\mu E\Big),
\label{batot}
\eeq
where vectors $\bar{a},...\bar{e}$ are
\[\bar{a}_\mu=p_\mu p_--n_\mu^-(\bt^2+p^2),\]
\[\bar{b}_\mu=2p_-r_{1\mu}+(p-2\bt)_\mu r_{1-}-n^-_\mu\Big(2(pr_1)+
q_2^2\frac{r_{1-}}{p_-}\Big),\]
\[\bar{c}_\mu=2p_-r_{2\mu}+(p-2\bt)_\mu r_{2-}-n^-_\mu\Big(2(pr_2)+
q_2^2\frac{r_{2-}}{p_-}\Big),\]
\[\bar{d}_\mu=0,\]
\[\bar{e}_\mu=2p_-n^+_{\mu}+(p-2\bt)_\mu-n^-_{\mu}\Big(2p_++
\frac{q_2^2}{p_-}\Big)\]
and the colour factor is
\beq
C_2=f^{da_2c}f^{db_2e}f^{eb_1a_1} \ .
\eeq

\subsection{Fig. \ref{fig3},C}
Here on the right we have the well-known Lipatov vertex
$
f^{a_1b_1d}L_{1\nu},
$
where the momentum part is
\beq
L_{1\nu}
=a_{1\nu}+b_1n^+_\nu+c_1n^-_\nu.
\label{lip}
\eeq
Here
\[a_1=q_1+r_1,\ \ b_1=\frac{r_1^2}{q_{1+}}-2r_{1-},\ \
c_1=\frac{q_1^2}{r_{1-}}-2q_{1+}.\]
On the left, however, we have a new vertex RP$\to$RP. Using the
effective action we find that it consists of two terms, the term coming
from the standard 4-gluon interaction $Z_1$ and the induced term $Z_2$.
Calculations give
\begin{equation}
Z^{(1)}_{\mu\nu}=ig^2 \Big[
f^{a_2ce}f^{b_2de}
\left( 2n^{+}_{\mu}n^{-}_{\nu} -n^{-}_{\mu}n^{+}_{\nu} -g_{\mu\nu} \right)
+
f^{a_2de}f^{b_2ce}
\left( 2n^{-}_{\mu}n^{+}_{\nu} -n^{+}_{\mu}n^{-}_{\nu} -g_{\mu\nu} \right)
\Big]
\label{ea2}
\end{equation}
and
\begin{equation}
Z^{(2)}_{\mu\nu}=ig^2 q_{2\perp}^2
\left(\frac{f^{a_2ce}f^{b_2de}}{-p_- r_{2-}}
+\frac{f^{a_2de}f^{b_2ce}}{t_{1-} r_{2-}}\right) n^{-}_{\mu}n^{-}_{\nu}
+ig^2 r_{2\perp}^2
\left(\frac{f^{a_2ce}f^{b_2de}}{-t_{1+} q_{2+}}
+\frac{f^{a_2de}f^{b_2ce}}{p_+ q_{2+}}\right) n^{+}_{\mu}n^{+}_{\nu}\ .
\label{ea4}
\end{equation}

Multiplying these terms by the Lipatov vertex from the right we
correspondingly get two terms
\[
{\cal B}^{(1)}_\mu=g^3f^{a_1b_1d}\frac{1}{t_1^2+i0}
\Big\{f^{a_2ce}f^{b_2de}
\Big[2a_{1-}n^+_\mu-a_{1+}n^-_\mu-a_{1\mu}\]\[
+n^+_\mu\Big(\frac{r_1^2}{q_{1+}}-2r_{1-}\Big)-2n^-_\mu
\Big(\frac{q_1^2}{r_{1-}}
-2q_{1+}\Big)\Big]\]\[+
f^{a_2de}f^{b_2ce}\Big[2a_{1+}n^-_\mu-a_{1-}n^+_\mu-a_{1\mu}-2n^+_\mu
\Big(\frac{r_1^2}{q_{1+}}-2r_{1-}\Big)+n^-_\mu
\Big(\frac{q_1^2}{r_{1-}}-2q_{1+}\Big)\Big]\Big\}
\]
and
\[
{\cal B}^{(2)}_\mu=
g^3f^{a_1b_1d}\frac{1}{t^2+i0}
\Big[n^-_\mu z_1\Big(\frac{r_1^2}{q_{1+}}+a_{1-}-2r_{1-}\Big)+
n^+_\mu z_2\Big(\frac{q_1^2}{r_{1-}}+a_{1+}-2q_{1+}\Big)\Big],
\]
where we denoted
\[
z_1\equiv  q_{2\perp}^2
\left(\frac{f^{a_2ce}f^{b_2de}}{-p_- r_{2-}}
+\frac{f^{a_2de}f^{b_2ce}}{t_{1-} r_{2-}}\right)
\]
and
\[
z_2\equiv
 r_{2\perp}^2
\left(\frac{f^{a_2ce}f^{b_2de}}{-t_{1+} q_{2+}}
+\frac{f^{a_2de}f^{b_2ce}}{p_+ q_{2+}}\right).
\]

In practice it is convenient to split the total contribution
into two parts having different colour factors. So the contribution
from Fig. \ref{fig3},C consists of two different amplitudes
\[
{\cal A}_{3\mu}=
g^3C_3\frac{1}{t_1^2+i0}
\Big[2a_{1-}n^+_\mu-a_{1+}n^-_\mu-a_{1\mu}
+n^+_\mu\Big(\frac{r_1^2}{q_{1+}}-2r_{1-}\Big)-2n^-_\mu
\Big(\frac{q_1^2}{r_{1-}}
-2q_{1+}\Big)\]\beq+
n^-_\mu \frac{q_2^2}{-p_-r_{2-}}\Big(\frac{r_1^2}{q_{1+}}+
a_{1-}-2r_{1-}\Big)+
n^+_\mu \frac{r_1^2}{-t_{1+}q_{2+}}\Big(\frac{q_1^2}{r_{1-}}+a_{1+}-2q_{1+}\Big)\Big]
\label{a3}
\eeq
and
\[
{\cal A}_{4\mu}=
g^3C_4\frac{1}{t_1^2+i0}
\Big[2a_{1+}n^-_\mu-a_{1-}n^+_\mu-a_{1\mu}
-2n^+_\mu\Big(\frac{r_1^2}{q_{1+}}-2r_{1-}\Big)+n^-_\mu
\Big(\frac{q_1^2}{r_{1-}}
-2q_{1+}\Big)\]\beq+
n^-_\mu \frac{q_2^2}{t_{1-}r_{2-}}\Big(\frac{r_1^2}{q_{1+}}
+a_{1-}-2r_{1-}\Big)+
n^+_\mu \frac{r_1^2}{p_+q_{2+}}\Big(\frac{q_1^2}{r_{1-}}+
a_{1+}-2q_{1+}\Big)\Big],
\label{a4}
\eeq
where the colour factors are
\beq
C_3=f^{a_1b_1d}f^{a_2ce}f^{b_2de}=-C_2
\eeq
and
\beq
C_4=f^{a_1b_1d}f^{a_2de}f^{b_2ce}=C_1 \ .
\eeq

\subsection{Fig. \ref{fig3},D}
The contribution from this diagram comes from two Lipatov vertices
$L_{1\nu_1}$ and $L_{2\nu_2}$
coupled to the triple gluon vertex
\[
\Gamma_{\nu_1\mu,\nu_2}(t_1,p,t_2)=
-gf^{d_1 c d_2}\Big((p+t_2)_{\nu_1}g_{\mu\nu_2}+
(t_1-t_2)_{\mu}g_{\nu_1\nu_2}+(-t_1-p)_{\nu_2}g_{\mu\nu_1}\Big),
\]
where $t_1=q_1-r_1$, $t_2=q_2-r_2$
The two Lipatov vertices are transversal
$(t_1L_1)=(t_2L_2)=0$.
So in $\Gamma_{\nu_1\mu\nu_2}$
we can drop terms with $t_{1\nu_1}$ and $t_{2\nu_2}$ and take
\beq
\Gamma_{\nu_1\mu\nu_2}=-f^{d_1cd_2}\Big(2t_{2\nu_1}g_{\mu\nu_2}+
(t_1-t_2)_\mu g_{\nu_1\nu_2}-2t_{1\nu_2}g_{\mu_1}\Big).
\eeq
As a result we find a compact expression for the contribution
from Fig. \ref{fig3},D
\beq
{\cal A}_{5\mu}=g^3C_5
\frac{1}{(t_1^2+i0)(t_2^2+i0)}
\Big(2(t_2L_1)L_{2\mu}-2(t_1L_2)L_{1\mu}
+(L_1L_2)(t_1-t_2)_\mu \Big).
\label{a5}
\eeq
Here the colour factor is
\beq
C_5=f^{a_1d_1b_1}f^{d_1 c d_2}f^{a_2d_2b_2}=C_1+C_2,
\eeq
where we have used the Jacoby identity.

Note that for some calculations a more explicit form may be
preferable, which is given in the Appendix.

\subsection {Symmetrization and particular configurations}
Contributions calculated above refer to a fixed order of upper and lower
reggeons
\[
{\cal A}(q_2,a_2;q_1,a_1|r_2,b_2;r_1,b_1)\equiv{\cal A}(2,1|2,1).
\]
The total amplitude is obtained after we sum it with the contributions
with interchange of upper and lower gluons
For ${\cal A}_i$ with $i=1,2,3,4$ each interchange gives a new diagram
so that
the total amplitude is
\[{\cal A}_i^{tot}={\cal A}_i(2,1|2,1)+{\cal A}_i(2,1|1,2)+
{\cal A}_i(1,2|2,1)+{\cal A}_i(1,2|1,2),\ \ i=1,2,3,4.\]
For $i=5$ simultaneous interchange of upper and lower reggeons does not give
a new contribution. So in this case
\[ {\cal A}_5^{tot}={\cal A}_5(2,1|2,1)+{\cal A}_5(2,1|1,2).\]

Each interchange combines interchange of momenta and colours.
In the general case this introduces a multitude of different
colour factors. To simplify we restrict ourselves with colour configurations
actually present in the applications. Inspecting Fig. \ref{fig1} we
see that the RR$\to$RRP vertex may appear in two different colour
configurations. One of them, Fig. \ref{fig1},A is diffractive
with respect to the targets but non-diffractive with respect to the
projectiles, D-ND configuration (of course there exists a similar
configuration with projectiles and targets reversed). The other configuration,
that of Fig. \ref{fig1},B is non-diffractive with respect to both
projectiles and targets, the ND-ND configuration.
In the D-ND configuration the general
colour coefficient $C(a_2,a_1|b_2,b_1)$ is to be convoluted with
$\delta_{b_1,b_2}$ . Then we obtain
\beq
C(a_2,a_1|b_2,b_1)\delta_{b_1b_2}=Nf^{a_2a_1c}\kappa^{D-ND},
\eeq
where $\kappa$ is just the number, different for different diagrams.
In the ND-ND configuration the general colour coefficient is to
be convoluted with $\delta_{a_1b_2}$ and we obtain
\beq
C(a_2,a_1|b_2,b_1)\delta_{a_1b_2}=Nf^{a_2b_1c}\kappa^{ND-ND},
\eeq
where again the number $\kappa$ is different for different diagrams.
The choice of convolution colours $a_1$ and $b_2$ is of course arbitrary
due to symmetry in both upper and lower reggeons.

Using these considerations for both configurations the total
amplitudes can be presented via their momentum parts as follows.
In the D-ND configuration
\[ {\cal A}_i^{tot}=Nf^{a_2a_1c}
=\kappa_i^{(1)}{\cal A}_i(q_2,q_1|r_2,r_1)+
\kappa_i^{(2)}{\cal A}_i(q_2,q_1|r_1,r_2)\]\beq+
\kappa_i^{(3)}{\cal A}_i(q_1,q_2|r_2,r_1)+
\kappa_i^{(4)}{\cal A}_i(q_1,q_2|r_1,r_2).
\label{tott}
\eeq
In the ND-ND configuration we have the same formula with
$f^{a_2a_1c}\to f^{a_2b_1c}$.

Simple calculations give for D-ND configuration
\[\kappa_i^{(1)}=\kappa_i^{(2)}=-\kappa_i^{(3)}=-\kappa_i^{(4)}\]
and
\[\kappa_1^{(1)}=\kappa_4^{(1)}=-\frac{1}{2},\ \
\kappa_2^{(1)}=-\kappa_3^{(1)}=1,\ \ \kappa_5^{(1)}=\frac{1}{2}\]
We recall that for $i=5$ in (\ref{tott}) only the first two terms
are to be taken into account.

For the ND-ND configuration coefficients $\kappa_i^{(k)}$ are given
in the table.
\begin{center}
\vspace*{0.5 cm}
{\bf Table of $\kappa_i^{(k)}$ for ND-ND configuration}
\vspace*{0.5 cm}

\begin{tabular}{|c|c|c|c|c|}
\hline
$k=$&1&2&3&4\\\hline
$\kappa_1$&1/2&0&1&1\\\hline
$\kappa_2$&-1&0&-1&-1/2\\\hline
\end{tabular}
\end{center}
Other coefficients are defined through $\kappa_1$ and $\kappa_2$
according to relations
\[ \kappa_3=-\kappa_2,\ \ \kappa_4=\kappa_1,\ \
 \kappa_5=\kappa_1+\kappa_2\]

\section{Transversality}
\subsection{Amplitudes ${\cal A}_i$, $i=1,...5$}
The obtained expressions for RR$\to$ RRP amplitudes are rather cumbersome.
A simple method to check transversality is just to calculate the product
$(p{\cal A})$ numerically. The corresponding calculations by a FORTRAN
program in both D-ND and ND-ND configurations show that the
constructed vertex is indeed transversal.
Nevertheless it is instructive to see this fact analytically.

We consider subsequently the five amplitudes ${\cal A}_i$,
$i=1,...5$ corresponding to contributions (\ref{atot}), (\ref{batot}),
(\ref{a3}), (\ref{a4}) and (\ref{a5}) respectively.

{\bf 1.} ${\cal A}_1$

Using orthogonality of $\bar{V}$ we find
\beq X_1=(p{\cal A}_1)=g^3C_1\frac{1}{t_1^2}
(\bar{A}t_+-\bar{B}q_{1+}-\bar{C}q_{2+}+\bar{E})
=g^3C_1\frac{1}{t_1^2}Z_1,
\label{x1}
\eeq
where $t_1=q_1-r_1$ and $\bar{A},...\bar{E}$ are given by (\ref{babcd}).
We find
\beq
Z_1=-3r_{1-}(q_{1+}+q_{2+})+(a_1,t+q_2)+q_1^2+q_2^2+2r_1^2-t_1^2
+r_1^2\frac{q_{2+}}{q_{1+}}-\frac{q_2^2r_1^2}{q_{1+}r_{1-}},
\label{z1}
\eeq
where $a_1=q_1+r_1$.

{\bf 2.} ${\cal A}_2$

Similarly to (\ref{x1}) we present
\beq
X_2=(p{\cal A}^{(2)})=
g^3C_2\frac{1}{t_1^2}(At_-+Br_{1-}+Cr_{2-}+E)=
g^3C_2\frac{1}{t_1^2}Z_2.
\label{x2}
\eeq
The explicit expressions for $A,...E$ are given in (\ref{abcd})
 We find
\beq
Z_2=-3q_{1+}(r_{1-}+r_{2-})-(a_1,\bar{t}-r_2)+r_1^2+r_2^2+2q_1^2-t_1^2
+q_1^2\frac{r_{2-}}{r_{1-}}-\frac{q_1^2r_2^2}{q_{1+}r_{1-}}
\label{z2}
\eeq

{\bf 3.} ${\cal A}_3$

We have
\[ p^\mu Z_{\mu\nu}^{(3)}=if^{a_2ce}f^{b_2de}
\Big[n_\nu^+\Big(2p_+-\frac{q^2}{r_{2-}}\Big)-n_\nu^-
\Big(p_-+\frac{r_2^2p_+}{t_{1+}q_{2+}}\Big)-p_\nu\Big].
\]
This has to  be multiplied by $L_{1\nu}$ given by (\ref{lip})
We get
\[ X_3=(p{\cal A}^{(3)})=g^3C_3\frac{1}{t_1^2}Z_3,\]
Calculations give
\[
Z_3=3q_{1+}p_--(a_1p)+r_1^2+r_2^2+2q_1^2
+2q_1^2\frac{r_{2-}}{r_{1-}}+q_2^2\frac{r_{1-}}{r_{2-}}
+r_1^2\frac{q_{2+}}{q_{1+}}+r_2^2\frac{q_{1+}}{q_{2+}}\]\beq-
\frac{q_2^2r_1^2}{q_{1+}r_{2-}}-\frac{q_1^2r_2^2}{q_{1+}r_{1-}}
-\frac{q_1^2r_2^2}{q_{2+}r_{1-}}.
\label{z3}
\eeq

{\bf 4.} ${\cal A}_4$

We have
\[ p^\mu Z_{\mu\nu}^{(4)}=if^{a_2de}f^{b_cde}
\Big[n_\nu^+\Big(2p_-+\frac{r^2}{q_{2+}}\Big)-n_\nu^-
\Big(p_+-\frac{q_2^2p_-}{t_{1-}r_{2-}}\Big)-p_\nu\Big].
\]
This again has to  be multiplied by (\ref{lip}).
We get
\[ X_4=(p{\cal A}^{(4)})=g^3C_4\frac{1}{t_1^2}Z_4,\]
where after simple calculations
\[
Z_4=3r_{1-}p_+-(a_1p)-q_1^2-q_2^2-2r_1^2
-2r_1^2\frac{q_{2+}}{q_{1+}}-r_2^2\frac{q_{1+}}{q_{2+}}
-q_1^2\frac{r_{2-}}{r_{1-}}-q_2^2\frac{r_{1-}}{r_{2-}}\]\beq+
\frac{q_1^2r_2^2}{q_{2+}r_{1-}}+\frac{q_2^2r_1^2}{q_{1+}r_{1-}}
+\frac{q_2^2r_1^2}{q_{1+}r_{2-}}.
\label{z4}
\eeq

{\bf 5.} ${\cal A}_5$

We again present
\beq
X_5=(p{\cal A}^{(4)})=-C_5\frac{1}{t_1^2}Z_5
\label{z5}
\eeq
and find
\[
Z_5=(L_1L_2)=
(a_1+b_1n^++c_1n^-,a_2+b_2n^++c_2n^-)\]\beq=
(a_1a_2)+b_1a_{2+}+b_2a_{1+}+c_1a_{2-}+c_2a_{1-}+b_1c_2+b_2c_1
\eeq
Note that this is not the total contribution from ${\cal A}^{(5)}$
but only half of it containing  $1/t_1^2$. The other half contains
$-1/t_2^2$ and so has a different structure from the other amplitudes.
It has to be taken into account in amplitudes with $1\lra 2$.
Also note the sign "-" in (\ref{z5}).
Calculations give
\beq
Z_5=(a_1a_2)-q_1^2\frac{r_{2-}}{r_{1-}}-q_2^2\frac{r_{1-}}{r_{2-}}-
r_1^2\frac{q_{2+}}{q_{1+}}-r_2^2\frac{q_{1+}}{q_{2+}}+
\frac{q_1^2r_2^2}{q_{2+}r_{1-}}+\frac{q_2^2r_1^2}{q_{1+}r_{2-}}.
\label{z5a}
\eeq

{\bf 6.} ${\cal A}^{tot}$

Summing our contributions we find
\beq
X^{tot}=(p{\cal A}^{tot})=g^3\frac{1}{t_1^2}(\sum_{i=1}^4C_iZ_i-
C_5Z_5)=g^3\frac{1}{t_1^2}\Big[ C_1(Z_1+Z_4-Z_5)+C_2(Z_2-Z_3-Z_5)\Big]
\eeq

\subsection{$Z_1+Z_4-Z_5$}
Since our expressions are long enough we separately consider terms
with double poles at $q_{i+}=0$ and $r_{i-}=0$, $i=1,2$, simple poles
and non-singular terms.

Suppressing the common factor $g^3/t_1^2$, the double pole contribution is
\[
Z^{(1+4-5)}_{dp}=
\Big(-\frac{q_2^2r_1^2}{q_{1+}r_{1-}}\Big)+
\Big(\frac{q_2^2r_1^2}{q_{1+}r_{1-}}
+\frac{q_2^2r_1^2}{q_{1+}r_{2-}}+
\frac{q_1^2r_2^2}{\qa\rb}\Big)
-\Big(\frac{q_1^2r_2^2}{q_{2+}r_{1-}}
+\frac{q_2^2r_1^2}{q_{1+}r_{2-}}\Big)=0,\]
where  the 3 terms in the brackets  correspond to contributions
from $Z_1,Z_4$ and $Z_5$. Obviously they cancel.

The single pole contribution is
\[
Z^{(1+4-5)}_{p}=
\Big(r_1^2\frac{q_{2+}}{q_{1+}}\Big)+
\Big(
-2r_1^2\frac{q_{2+}}{q_{1+}}-r_2^2\frac{q_{1+}}{q_{2+}}
-q_1^2\frac{r_{2-}}{r_{1-}}-q_2^2\frac{r_{1-}}{r_{2-}}\Big)
-\Big(
-q_1^2\frac{r_{2-}}{r_{1-}}-q_2^2\frac{r_{1-}}{r_{2-}}-
r_1^2\frac{q_{2+}}{q_{1+}}-r_2^2\frac{q_{1+}}{q_{2+}}\Big)=0.\]
Again the three terms correspond to contributions from $Z_1,Z_4,Z_5$

The three terms with the nonsingular contributions are
\[ Z^{(1+4-5)}_{ns}=
\Big(-3r_{1-}(q_{1+}+q_{2+})+(a_1,q_1+2q_2-r_1)+
q_1^2+q_2^2+2r_1^2-t_1^2\Big)\]
\[
+\Big(3r_{1-}(q_{1+}+q_{2+})-(a_1,q_1+q_2-r_1-r_2)
-q_1^2-q_2^2-2r_1^2\Big)-(a_1a_2).\]
We find for vectors multiplying $a_1$ in the first two terms
\[ q_1+2q_2-r_1-q_1-q_2+r_1+r_2=a_2\]
So in the sum all terms cancel except containing $t_1^2$:
$Z^{(1+4-5)}_{ns}=-t_1^2$.
So restoring the suppressed factor
$X^{(1+4-5)}=-g^3C_1$.
This means that the sum of all diagrams considered above is not
transversal by itself. Violation of transversality comes from the
contribution ${\cal A}_1$.

\subsection{$Z_2-Z_3-Z_5$}
In a similar fashion here we find the double pole contribution
\[
Z^{(2+3-5)}_{dp}=
-\Big(\frac{q_1^2r_2^2}{q_{1+}r_{1-}}\Big)
+\Big(\frac{q_2^2r_1^2}{q_{1+}r_{2-}}+\frac{q_1^2r_2^2}{q_{1+}r_{1-}}
+\frac{q_1^2r_2^2}{q_{2+}r_{1-}}\Big)
-\Big(\frac{q_1^2r_2^2}{q_{2+}r_{1-}}+
\frac{q_2^2r_1^2}{q_{1+}r_{2-}}\Big)=0.\]
The 3 terms in the brackets  correspond to contributions
from $Z_2,Z_3$ and $Z_5$. They cancel in the sum.

The single pole contribution is
\[
Z^{(2-3-5)}_{p}=
+\Big(q_1^2\frac{r_{2-}}{r_{1-}}\Big)
-\Big(2q_1^2\frac{r_{2-}}{r_{1-}}+q_2^2\frac{r_{1-}}{r_{2-}}
+r_1^2\frac{q_{2+}}{q_{1+}}+r_2^2\frac{q_{1+}}{q_{2+}}\Big)\]\[
-\Big(
-q_1^2\frac{r_{2-}}{r_{1-}}-q_2^2\frac{r_{1-}}{r_{2-}}-
r_1^2\frac{q_{2+}}{q_{1+}}-r_2^2\frac{q_{1+}}{q_{2+}}\Big).\]
They also give zero in the sum.

Finally the nonsingular contribution is
\[ Z^{(2-3-5)}_{ns}=
+\Big(-3q_{1+}(r_{1-}+r_{2-})-(a_1,q_1-r_1-2r_2)
+r_1^2+r_2^2+2q_1^2-t_1^2\Big)\]\[
-\Big(-3q_{1+}(r_{1-}+r_{2-})-(a_1,q_1+q_2-r_1-r_2)+r_1^2+r_2^2+
2q_1^2\Big)-(a_1a_2).\]
We find for vectors multiplying $a_1$ in the first 2 terms
\[-(q_1-r_1-2r_2-q_1-q_2+r_1+r_2)=a_2.\]
In the sum all terms cancel again except containing $t_1^2$:
$Z^{(2-3-5)}_{ns}=-t_1^2$.
So restoring the  factor in front
$X^{(2-3-5)}=-g^3 C_2$.
Again the sum of all diagrams considered above is not
transversal by itself. In this part violation of transversality
comes from contributions ${\cal A}_2$.

For the sum of all diagrams with fixed reggeon momenta we find
\beq
X^{tot}=-g^3(C_1+C_2)=-g^3C_5.
\label{xtot}
\eeq
This expression
is  antisymmetric under interchange
$(a_2,a_1|b_2,b_1)\lra(a_1,a_2|b_1,b_2)$ and does
not depend on the momenta of the 4 reggeons.
So it will vanish after
symmetrization in the sum ${\cal A}(2,1|2,1)+{\cal A}(1,2|1,2)$.
Thus after symmetrization
we find transversality of the constructed RR$\to$RRP vertex.

\section{On-mass-shell amplitudes}
On mass shell, at $p^2=0$, the physical amplitudes can be presented
via the physical polarization vector $\ep_\mu$, which we choose with the
properties
\beq
(p\ep)=(l\ep)=0,\ \ \ep_+=0,\ \ \ep_-=-\frac{(p\ep)_\perp}{p_+}.
\eeq
So the product with any vector $v$
\beq
(v\ep)=(v\ep)_\perp-\frac{v_+}{p_+}(p\ep)_\perp.
\eeq

Amplitudes ${\cal A}_i$, $i=1,...,5$ take the following form
on the mass shell multiplied by the polarization vector $\ep$

{\bf 1.} ${\cal A}_1$

Obviously it is sufficient to transform our coefficients $a,b,...,e$.
We have
\[
a_{\ep}\equiv(a\ep)=0,
\]\[
b_{\ep}\equiv(b\ep)=2p_+(q_1\ep)+q_{1+}(p-2t,\ep).\]
Since $p-2t=-p-2r_2$, $(p-2t,\ep)=-2(r_2\ep)$, which gives
\[b_{\ep}=2p_+\Big((q_1\ep)_\perp-\frac{q_{1+}}{p_+}(p\ep)_\perp\Big)-
2q_{1+}(r_2\ep)_\perp
=2p_+(q_1\ep)_\perp - 2q_{1+} (p+r_2,\ep)_\perp.\]
Similarly,
\[c_{\ep}=2p_+\Big((q_2\ep)_\perp-\frac{q_{2+}}{p_+}(p\ep)_\perp\Big)-
2q_{2+}(r_2\ep)_\perp
=2p_+(q_2\ep)_\perp - 2q_{2+} (p+r_2,\ep)_\perp.\]
and finally
\[e_{\ep}=-2(r_2\ep)_\perp-2(p\ep)_\perp.\]

{\bf 2.} ${\cal A}_2$

Again we transform coefficients $\bar{a},...,\bar{e}$.
We have
\[\bar{a}_\ep=(p\ep)_\perp\frac{t^2}{p_+},\]
\[\bar{b}_\ep=2p_-(r_1\ep)+r_{1-}(p-2t,\ep)+
\frac{(p\ep)_\perp}{p_+}\Big(2(pr_1)+q_2^2\frac{r_{1-}}{p_-}\Big),\]
where $t=p-q_2$, so that $(p-2t,\ep)=2(q_2\ep)$. We find
\[\bar{b}_\ep=2p_-(r_1\ep)_\perp+2r_{1-}\Big((q_2\ep)_\perp-
(p\ep)_\perp\frac{q_{2+}}{p_+}\Big)+2(p\ep)_\perp\Big(r_{1-}-r_{1-}
\frac{q_2^2}{p_\perp^2}+\frac{(pr_1)_\perp}{p_+}\Big).\]
Similarly
\[\bar{c}_\ep=2p_-(r_2\ep)_\perp+2r_{2-}\Big((q_2\ep)_\perp-
(p\ep)_\perp\frac{q_{2+}}{p_+}\Big)+2(p\ep)_\perp\Big(r_{2-}-r_{2-}
\frac{q_2^2}{p_\perp^2}+\frac{(pr_2)_\perp}{p_+}\Big).\]
Finally
\[\bar{e}_\ep=2(q_2\ep)_\perp+2(p\ep)_\perp\Big(1-\frac{q_{2+}}{p_+}
-\frac{q_2^2}{p_\perp^2}\Big).\]

{\bf 3.} ${\cal A}_3$

We present
\beq
{\cal A}_{3\ep}=g^3C_3\frac{1}{t_1^2}B_3,
\eeq
where
\[
B_3=2(p\ep)_\perp\frac{a_{1+}}{p_+}-(a_1\ep)_\perp
+2\frac{(p\ep)_\perp}{p_+}\Big(\frac{q_1^2}{r_{1-}}-2q_{1+}\Big)-
2\frac{(p\ep)_\perp q_2^2}{p_\perp^2}\Big(\frac{r_1^2}{q_{1+}}
-r_{1-}\Big).\]
Somewhat transforming we find
\[
B_3=-(a_1\ep)_\perp+2(p\ep)_\perp\Big[-\frac{q_{1+}}{p_+}+\frac{q_1^2}{p_+r_{1-}}
-\frac{q_2^2}{p_\perp^2 r_{2-}}\Big(\frac{r_1^2}{q_{1+}}-r_{1-}\Big)
\Big],
\]
where $a_1=q_1+r_1$

{\bf 4.} ${\cal A}_4$

We present
\beq
{\cal A}_{4\ep}=g^3C_4\frac{1}{t_1^2}B_4,
\eeq
where
\[
B_4=-(p\ep)_\perp\frac{q_{1+}}{p_+}-(a_1\ep)_\perp
-\frac{(p\ep)_\perp}{p_+}\Big(\frac{q_1^2}{r_{1-}}-2q_{1+}\Big)
+\frac{(p\ep)_\perp q_2^2}{p_+r_{1-}r_{2-}}\Big(\frac{r_1^2}{q_{1+}}
-r_{1-}\Big).\]
Again transforming to obtain final expresion
\[
B_4=-(a_1\ep)_\perp+\frac{(p\ep)_\perp}{p_+}\Big(q_{1+}-
\frac{q_1^2}{r_{1-}}
-\frac{q_2^2}{ r_{2-}}+\frac{q_2^2r_1^2}{q_{1+}r_{1-}r_{2-}}\Big).
\]

{\bf 5.} ${\cal A}_5$

Multiplying by $\ep$ we find
\[L_{1\ep}=(L_1\ep)=
(a_1\ep)_\perp-\frac{(p\ep)_\perp}{p_+}
\Big(\frac{q_1^2}{r_{1-}}-q_{1+}\Big),
\]
\[L_{2\ep}=(L_2\ep)=
(a_2\ep)_\perp-\frac{(p\ep)_\perp}{p_+}
\Big(\frac{q_2^2}{r_{2-}}-q_{2+}\Big)\ 
\]
and finally
\[
(t_1-t_2)_\ep=(t_1-t_2,\ep)=
(t_1-t_2,\ep)_\perp-\frac{(p\ep)_\perp}{p_+}(q_{1+}-q_{2+}).
\]
The product $({\cal A}_5\ep)$ will be given by Eq. (\ref{a5}) with vectors
substituted by their products with $\ep$ given above.

\section{Asymptotics for large $q_{1+}$ or $r_{1-}$ with fixed $p$}
For applications the behaviour of the vertex
at large values of longitudinal momenta has the utter importance,
since one has to integrate over them
when the vertex is inserted into the amplitude. The necessary condition for the possibility of
this integration is that the amplitude should vanish at high values of
longitudinal momenta. Note that
in the inclusive
cross-section momentum $p$ of the observed gluon is fixed. This means
that sums $\qa+\qb$ and $\ra+\rb$ remain finite when one of the
longitudinal momenta tends to infinity. Having in mind that in the D-ND
configuration upper and lower reggeons enter in the different manner
we have to study separately cases of $\qa\to \infty$ and $\ra\to\infty$.

\subsection{$q_{1+}\to\infty$, $p$ fixed}
We consider subsequently our amplitudes ${\cal A}_i$, $i=1,...5$.

{\bf 1.} ${\cal A}_1$

Here $t=q_1+q_2-r_1$. Since $q_2=p-q_1$, $t$ is finite.
Of the two denominators one is finite the other grows as $q_{1+}$.
So non-vanishing terms come from the ones in the numerator which grow as
$q_{1+}$ or faster.
 Inspecting coefficients $a,b,c,e$ we conclude: $a$ and $e$ are finite,
 \[b=2p_+q_1+q_{1+}(p-2t)-q_{1+}n^+\Big(2p_-+\frac{r_2^2}{p_+}\Big),\]
 \[c=2p_+q_2+q_{2+}(p-2t)-q_{2+}n^+\Big(2p_-+\frac{r_2^2}{p_+}\Big).\]
Turning to $\bar{A},...\bar{E}$ we find that
$\bar{A},\bar{B}$ are finite with $\bar{B}=\bar{C}$. So the contribution
$a\bar{A}-b\bar{B}-c\bar{C}$ is finite. We are left with only $e\bar{E}$.

At large $q_{1+}$
$\bar{E}=\bar{E}_0-2q_{2+}r_{1-}$,
where
\[\bar{E}_0=(r_1+q_1,t+q_2)+q_2^2-q_1^2+(q_1-r_1)^2+2r_{1-}q_{1+}=
r_{1-}q_{2+}-r_{1-}q_{1+}\]
We find
$\bar{E}=-r_{1-}(q_{1+}+q_{2+})$
and is finite. So there are no growing terms in the numerator and
in the limit $q_{1+}\to\infty$ ${\cal A}_1=0$.

{\bf 2.} ${\cal A}_2$

Here $\bar{t}=q_1-r_1-r_2$ and grows as $q_{1+}$. The two denominators
both grow as $q_{1+}$. Possible terms non-vanishing at $q_{1+}\to\infty$
may come from the ones in the numerator growing as $q_{1+}^2$ or faster.

Coefficients $\bar{a},...\bar{e}$ are in this limit
\[\bar{a}=n^-\bar{t}^2,\ \ \bar{b}=-2\bar{t}r_{1-},\ \ \bar{c}=-3\bar{t}r_{2-},
\ \ \bar{e}=-2\bar{t}.\]
Terms $A,...E$ are
\[A=3q_{1+},\ \ B=C=4q_{1+},\ \ E=E_0+2q_{1+}(\bar{t}_--r_{2-}),\]
where
\[E_0=-(q_1+r_1,\bar{t}-r_2)+r_2^2-r_1^2+(q_1-r_1)^2+2q_{1+}r_{1-}=
-q_{1+}(\bar{t}-r_2)_--q_{1+}r_{1-}.\]
So
\[E=q_{1+}(\bar{t}-r_{2-}-r_{1-})=-2q_{1+}(r_{1-}+r_{2-}).\]
Thus
\[\bar{a}A+\bar{b}B+\bar{c}C+\bar{e}E=
6q_{1+}^2(r_{1-}+r_{2-})-2q_{1+}^2r_{1-}-2q_{1+}^2r_{2-}+
4q_{1+}^2(r_{1-}+r_{2-})=2q_{1+}(r_{1-}+r_{2-}).\]
The denominator is
$\bar{t}^2t_1^2=4q_{1+}^2r_{1-}(r_{1-}+r_{2-})$.
Thus in the limit $q_{1+}\to \infty$
\beq
{\cal A}_{2+}=-g^3C_2 \frac{1}{2r_{1-}}.
\label {a2q}
\eeq

{\bf 3.} ${\cal A}_3$

In the limit $\qa\to\infty$ the square bracket in (\ref{a3}) is
$-a_+n^--a+4n^-\qa.$
So the growing "+" component is just $2\qa$ and
in the limit
\beq
{\cal A}_{3+}=-g^3C_3\frac{1}{\ra}.
\label{a3q}
\eeq

{\bf 4.} ${\cal A}_4$

In the limit $\qa\to\infty$ the square bracket in (\ref{a4})is
$2\qa n^--a+n^-\qa.$
So the growing "+" component is just $-\qa$
and in this limit
\beq
{\cal A}_{4+}=g^3C_4\frac{1}{2\ra}.
\label{a4q}
\eeq

{\bf 5.} ${\cal A}_5$

The two denominators grow as $\qa$ each. So we have to search for
terms in the numerator which grow as $\qa^2$ or faster.
We have
\[
(t_2L_1)=(q_2-r_2,q_1+r_1)+b_1\qb-c_1\rb\]\[=
\qb\ra-\qa\rb-2\qb\ra+2\qa\rb=\qa\rb-\qb\ra.
\]
We also have in the limit $\qa\to\infty$ $ L^{(2)}=a_2-2n^-\qa$,
so that the growing "+" component is
$L_{2+}=-\qa$. As a result
\[L_{2+}(t_2L_1)-(1\lra 2)=
\qa(\qb\ra-\qa\rb)-\qb(\qa\rb-\qb\ra)\]\[=
\qa(\qb\ra-\qa\rb+\qa\rb-\qb\ra)=0.\]

So the terms growing with $\qa$ come from the third term in (\ref{a5}).
We find in the limit $\qa\to\infty$
\[(L_1L_2)=
(a_1a_2)+b_1a_{2+}+b_2a_{1+}+c_1a_{2-}+c_2a_{1-}
+b_1c_2+b_2c_1\]\[=
\qa\rb+\qb\ra-2\qa\rb-2\qb\ra-2\qa\rb-2\qb\ra+4\qa\rb+4\qb\ra\]\[=
\qa\rb+\qb\ra.\]
So we find the "+"component
\[(t_{1+}-t_{2+})(L^{(1)}L^{(2)})=2\qa(\qa\rb+\qb\ra)
=2\qa^2(\rb-\ra)\]
and thus
\beq
{\cal A}_{5+}=g^3C_5\Big(\frac{1}{2\ra}
-\frac{1}{2\rb}\Big).
\label{a5q}
\eeq

{\bf 6.}

After addition of the symmetrized contributions we find
for "+" components suppressing the common factor $g^3f^{a_2a_1c}$
or $g^3f^{a_2b_1c}$ for D-ND and ND-ND configurations respectively:
\[
{\cal A}_2^{tot}=-(\kappa_2^{(1)}-\kappa_2^{(3)})\frac{1}{2r_{1-}}-
(\kappa_2^{(2)}-\kappa_2^{(4)})\frac{1}{2r_{2-}},\ \
{\cal A}_3^{tot}=-(\kappa_3^{(1)}-\kappa_3^{(3)})\frac{1}{r_{1-}}-
(\kappa_3^{(2)}-\kappa_3^{(4)})\frac{1}{r_{2-}},\]
\[
{\cal A}_4^{tot}=(\kappa_4^{(1)}-\kappa_4^{(3)})\frac{1}{2r_{1-}}+
(\kappa_2^{(2)}-\kappa_2^{(4)})\frac{1}{2r_{2-}},\ \
{\cal A}_5^{tot}=\frac{1}{2}\Big(\kappa_5^{(1)}-\kappa_5^{(3)}\Big)
\Big(\frac{1}{\ra}-\frac{1}{\rb}\Big).\]

For D-ND configuration we have shown
\[ \kappa_2^{(1)}=\kappa_2^{(3)}=-\kappa_2^{(2)}=-\kappa_2^{(4)}=1,\ \
\kappa_3^{(1)}=\kappa_3^{(3)}=-\kappa_3^{(2)}=-\kappa_3^{(4)}=-1,\]
\[ \kappa_4^{(1)}=\kappa_4^{(3)}=-\kappa_4^{(2)}
=-\kappa_4^{(4)}=-\frac{1}{2},\ \
 \kappa_5^{(1)}=\kappa_5^{(3)}=-\kappa_5^{(2)}=-\kappa_5^{(4)}
 =\frac{1}{2}.\]
So all contributions are zero.
This means that each of the diagrams studied above separately behaves as
$1/\qa$ as $\qa\to\infty$.

For ND-ND configuration, using $\kappa_i^{(k)}$ from the table we find
\[
{\cal A}_2^{tot}=\Big(\frac{1}{2}+\frac{1}{2}\Big)\frac{1}{\ra}+
\frac{1}{4}\frac{1}{\rb},\ \
{\cal A}_3^{tot}=\Big(-1-1\Big)\frac{1}{\ra}
-\frac{1}{2}\frac{1}{\rb},\]
\[
{\cal A}_4^{tot}=\Big(\frac{1}{4}+\frac{1}{2}\Big)\frac{1}{\ra}+
\frac{1}{2}\frac{1}{\rb},\ \
{\cal A}_5^{tot}=\frac{1}{4}\frac{1}{\ra}-
\frac{1}{4}\frac{1}{\rb}.\]
Subsequent terms on the right-hand side correspond to contributions from
$i=1,3$ and 4.
We observe that in this case separate contributions do not vanish in
the limit $\qa\to \infty$. However, the sum of them does vanish in this
limit.

\subsection{$\ra\to\infty$, $p$ fixed}
Again we subsequently study contributions ${\cal A}_i$, $i=1,...5$.

{\bf 1.} ${\cal A}_1$

In this case $t=q_1+q_2-r_1$ grows and the two denominators each grow
as $\ra$. So we have to separate terms in the numerator growing as
$\ra^2$.

Coefficients $a,...e$ are at large $t$
\[a=-n^+t^2,\ \ b=-2t\qa,\ \ c=-2t\qb,\ \ e=-2t.\]
Terms $\bar{A},...\bar{E}$ are
\[\bar{A}=3t_-,\ \ \bar{B}=\bar{C}=4t_-,\ \
\bar{E}=\bar{E}_0-2\ra(t_++\qb),\]
where
\[\bar{E}_0=(r_1+q_1,t+q_2)-(r_1-q_1)^2-2\ra\qa=
\ra(t_++\qb)-\ra\qa.\]
As a result
\[\bar{E}=-\ra(t_++\qb)-\ra\qa=-2\ra(\qa+\qb).\]

Thus the growing "-" component is
\[a_-\bar{A}-b_-\bar{B}-c_-\bar{C}+e_-\bar{E}=
-3t^2t_- +8t_-^2\qa+8t_-^2\qb-4t_-^2(\qa+\qb)=
=-2t_-^2(\qa+\qb).\]
The denominator is $4t_-^2\qa(\qa+\qb)$ so that finally
\beq
{\cal A}_{1-}=g^3C_1\frac{1}{2\qa}.
\label{a1r}
\eeq

{\bf 2.} ${\cal A}_2$

Here $\bar{t}=q_1-r_1-r_2$ and is finite.
As a result coefficients $\bar{a}$ and $\bar{e}$ are finite.
The rest
\[\bar{b}=2p_-r_1+\ra(p-2\bar{t})-n^-\ra\Big(2p_++\frac{q_2^2}{p_-}\Big),\]
\[\bar{c}=2p_-r_2+\rb(p-2\bar{t})-n^-\rb\Big(2p_++\frac{q_2^2}{p_-}\Big).\]
Terms $A,...E$ are
\[ A=3\qa,\ \ B=C=4\qa,\ \
 E=E_0+2\qa(\bar{t}_--\rb),\]
where
\[E_0=-(q_1+r_1)(\bar{t}-r_2)-(q_1-r_1)^2-2\qa\ra=
-\qa(\bar{t}_--\rb)-\qa\ra.\]
So
\[E=\qa(\bar{t}_--\rb)-\qa\ra=2\qa p_-\]
and is finite. We observe that all growing terms cancel and
in the limit $\ra\to\infty$ ${\cal A}_2=0$.

{\bf 3.} ${\cal A}_3$

In the limit $\ra\to\infty$ the square bracket in (\ref{a3}) is
$2\ra n^+-r_1-2n^+\ra$.
So the growing "-" component is  $-\ra$
and in this limit
\beq
{\cal A}_{3-}=g^3C_3\frac{1}{2\qa}.
\label{a3r}
\eeq

{\bf 4.} ${\cal A}_4$

In the limit $\ra\to\infty$ the square bracket in (\ref{a3}) is
$-\ra n^+-a+4n^+\ra$
So the growing "-" component is  $2\ra$
and in this limit
\beq
{\cal A}_{4-}=-g^3C_4\frac{1}{\qa}.
\label{a4r}
\eeq

{\bf 5.} ${\cal A}_5$

The two denominators grow as $\ra$ each. So we have to search for
terms in the numerators which grow as $\ra^2$ or faster.
We have seen that
\[
(t_2L_1)=(q_2-r_2,q_1+r_1)+b_1\qb-c_1\rb\]\[=
\qb\ra-\qa\rb-2\qb\ra+2\qa\rb=\qa\rb-\qb\ra.
\]
We also have in the limit $\ra\to\infty$ $ L_2=r_2-2n^+\ra$,
so that the growing "-" component is
$L_+^{(2)}=-\ra$. As a result
\[L_{2-}(t_2L_1)-(1\lra 2)=
\ra(\qb\ra-\qa\rb)-\rb(\qa\rb-\qb\ra)\]\[=
\ra(\qb\ra-\qa\rb+\qa\rb-\qb\ra)=0.\]

So the terms growing with $\ra$ come again from the third term in (\ref{a5}).
We find in the limit $\ra\to\infty$
\[(L_1L_2)=
(a_1a_2)+b_1a_{2+}+b_2a_{1+}+c_1a_{2-}+c_2a_{1-}
+b_1c_2+b_2c_1\]\[=
\qa\rb+\qb\ra-2\qa\rb-2\qb\ra-2\qa\rb-2\qb\ra+4\qa\rb+4\qb\ra\]\[=
\qa\rb+\qb\ra\]
So we find the "-"component
\[(t_{1-}-t_{2-})(L^{(1)}L^{(2)})=-2\ra(\qa\rb+\qb\ra)
=2\ra^2(\qa-\qb)\]
and as a result
\beq
{\cal A}^{(5)}_-=g^3C_5\Big(\frac{1}{2\qa}
-\frac{1}{8\qb}\Big).
\label{a5r}
\eeq

{\bf 6.}

After addition of the symmetrized contributions and suppressing the
common factors as before we find for "-" components
\[
{\cal A}_1^{tot}=(\kappa_1^{(1)}+\kappa_1^{(2)})\frac{1}{2\qa}+
(\kappa_1^{(3)}+\kappa_1^{(4)})\frac{1}{2\qb},\ \
{\cal A}_3^{tot}=(\kappa_3^{(1)}+\kappa_3^{(2)})\frac{1}{2\qa}+
(\kappa_3^{(3)}+\kappa_3^{(4)})\frac{1}{2\qb},\]
\[
{\cal A}_4^{tot}=-\Big((\kappa_4^{(1)}+\kappa_4^{(2)})\frac{1}{\qa}-
(\kappa_4^{(3)}+\kappa_4^{(4)})\frac{1}{\qb}\Big),\ \
{\cal A}_5^{tot}=\frac{1}{2}(\kappa_5^{(1)}+\kappa_5^{(2)})
\Big(\frac{1}{\qa}-\frac{1}{\qb}\Big).\]

For D-ND configuration
\[\kappa_1^{(1)}+\kappa_1^{(2)}=-1,\ \ \kappa_3^{(1)}+\kappa_3^{(2)}=-2,\ \
\kappa_4^{(1)}+\kappa_4^{(2)}=-1,\ \ \kappa_5^{(1)}+\kappa_5^{(2)}=1\]
and for all $i$
\[\kappa_i^{(3)}+\kappa_i^{(4)}=-\kappa_i^{(1)}-\kappa_i^{(2)}\]
So we get
\[
{\cal A}_1^{tot}=-\frac{1}{2}\Big(\frac{1}{\qa}-\frac{1}{\qb}\Big),\ \
{\cal A}_3^{tot}=-\Big(\frac{1}{\qa}-\frac{1}{\qb}\Big),\]
\[
{\cal A}_4^{tot}=\Big(\frac{1}{\qa}-\frac{1}{\qb}\Big),\ \
{\cal A}_5^{tot}=\frac{1}{2}\Big(\frac{1}{\qa}-\frac{1}{\qb}\Big).\]
So unlike the limit $\qa\to\infty$ in this configuration the
individual contributions do not vanish in the limit $\ra\to\infty$.
However, their sum vanishes.

For ND-ND configuration we find using the Table
\[
{\cal A}_1^{tot}=\frac{1}{2}\Big(\frac{1}{2}\,\frac{1}{\qa}
+2\frac{1}{\qb}\Big),\ \
{\cal A}_3^{tot}=\frac{1}{2}\Big(\frac{1}{\qa}+
\frac{3}{2}\,\frac{1}{\qb}\Big),\]
\[
{\cal A}_4^{tot}=-\Big(\frac{1}{2}\,\frac{1}{\qa}
+2\frac{1}{\qb}\Big),\ \
{\cal A}_5^{tot}=-\frac{1}{4}\Big(\frac{1}{\qa}-\frac{1}{\qb}\Big)\]
The individual contributions do not vanish again. Their sum is
\[
{\cal A}^{tot}=\sum_{i=1}^5{\cal A}_i^{tot}=
\frac{1}{\qa}\Big(\frac{1}{4}+\frac{1}{2}-\frac{1}{2}-\frac{1}{4}\Big)
+\frac{1}{\qb}\Big(1+\frac{3}{4}-2+\frac{1}{4}\Big)
=0\]
So it vanishes at $\ra\to\infty$.

Note that at $\ra\to \infty$ the leading contribution comes from the
"-" component of the vertex. So one may expect still better convergence
for the amplitude on the mass shell multiplied by the polarization
vector $\ep$ with a zero "+" component. Numerical calculations show that
this is indeed so in the D-ND configuration due to cancellations
in the sum with interchanged $\ra$ and $\rb$. In this case the vertex
behaves as $1/\ra^2$ at $\ra\to \infty$. However, the same numerical
calculations show that this result does not hold for the ND-ND
configuration, in which the amplitude behaves as $1/\ra$.

\section{Poles in longitudinal momenta}
Here we present contributions with poles at $\qa,\qb,\ra,\rb=0$
coming from the induced vertices in the effective action formalism.
In the case of gluon production in the collision of a single projectile
on several targets these poles cancel with the singularities coming from
rescattering contributions ~\cite{ref1,ref2,ref3}. In our case there
is no rescattering and one might think that these poles cancel in the
total amplitude after taking into account all permutations of
interacting reggeons. This possibility was advocated in ~\cite{odderon}
for the  second order odderon kernel.
However, we shall see that in our case pole singularities do not cancel
and remain in the total production amplitude. For applications this means
that one has to fix somehow the way to do longitudinal integrations in
presence of these poles. The requirement of the hermiticity
of effective action and  the structure of the simple reggeon
exchange prompt using integrations in the principal value sense.

Due to the complicated form of the production amplitudes the
simplest way to see existence of pole singularities at
$\qa,\qb,\ra\rb=0$ is by a numerical check. It indeed shows that in
both N-ND and ND-ND configurations the production amplitude contains
pole singularities at each $\qa,\qb,\ra$ and $\rb$ equal to zero
and also double pole singularities at, say, $\qa=\ra=0$.

However, it is instructive to extract pole contributions in the
analytical form to see their character and understand why they
cannot cancel. We shall subsequently consider pole contributions from
amplitudes ${\cal A}_i$ $i=1,...5$.

{\bf 1.} ${\cal A}_1$

Coefficients $a,...e$ are not singular. Singular terms in $\bar{A},...
\bar{E}$ are
\[\bar{A}=\frac{r_1^2}{q_{1+}},\ \ \bar{B}=0,\ \
\bar{C}=2\frac{r_1^2}{q_{1+}},\ \
\bar{E}=2\frac{r_1^2q_{2+}}{q_{1+}}-\frac{q_2^2r_1^2}{q_{1+}r_{1-}}.\]
Here $t=q_1+q_2-r_1$
Presenting the pole part of ${\cal A}_1$ as
\beq
{\cal A}_1=-g^3C_1\frac{1}{t^2t_1^2}X_1
\eeq
we find after trivial calculations
\[
X_{1\mu}=\frac{r_1^2}{q_{1+}}\Big[p_+(p_\mu-4q_{2\mu})+
n^+_\mu(4(pq_2)-4p_-q_{2+}-t^2-p^2)+4p_+q_{2+}n^-_\mu\Big]\]\beq
-\frac{q_2^2r_1^2}{q_{1+}r_{1-}}\Big[(p-2t)_\mu-
n^+_\mu\Big(2p_-+\frac{r_2^2}{p_+}\Big)+2p_+n^-_\mu\Big]
\eeq
On mass shell, multiplied by the polarization vector, it gives
\beq X_{1\ep}=-4(q_2\ep)_\perp\frac{r_1^2q_{2+}}{q_{1+}}
+2(p+r_2,\ep)_\perp\frac{q_2^2r_1^2}{q_{1+}r_{1-}}.
\eeq

{\bf 2.} ${\cal A}_2$

Coefficients $\bar{a},...\bar{e}$ are not singular.
Singular terms in $A,...E$ are
\[A=-\frac{q_1^2}{r_{1-}},\ \ B=0,\ \ C=-2\frac{q_1^2}{r_{1-}},\ \
E=2r_{2-}\frac{q_1^2}{r_{1-}}-\frac{q_1^2r_2^2}{q_{1+}r_{1-}}.\]
Here $\bar{t}=q_1-r_1-r_2$. We present
\beq
{\cal A}_2=-g^3C_2\frac{1}{\bar{t}^2t_1^2}X_2.
\eeq
Calculations give
\[
X_{2\mu}=\frac{q_1^2}{r_{1-}}\Big[ -p_-(p_\mu+r_{2\mu})+
4p_-r_{2-}n^+_\mu+n^-_\mu(\bar{t}^2+p^2+4(pr_2)-4p_+r_{2-}\Big]
\]\beq
 -\frac{q_1^2r_2^2}{q_{1+}r_{1-}}\Big[ (p-2\bar{t})_\mu+2p_-n^+_\mu
 -n^-_\mu\Big(2p_++\frac{q_2^2}{p_-}\Big)\Big].
\eeq
On mass shell, multiplied by the polarization vector,
\[
X_{2\ep}=-\frac{q_1^2}{r_{1-}}\Big[(p\ep)_\perp\Big(2r_{2-}+
\frac{2(p,2r_2-q_2)_\perp+q_2^2}{p_+}\Big)+4p_-(r_2\ep)_\perp\Big]
\]\beq
-2\frac{q_1^2r_2^2}{q_{1+}r_{1-}}\Big[(q_2\ep)_\perp+
(p\ep)_\perp\Big(\frac{q_{1+}}{p_+}-\frac{q_2^2}{p_\perp^2}\Big)\Big].
\eeq

{\bf 3.} ${\cal A}_3$

As before we present
\beq
{\cal A}_3=g^3C_3\frac{1}{t_1^2}X_3.
\eeq
We find
\beq
X_{3\mu}=n^+_\mu\Big(\frac{r_1^2}{q_{1+}}+\frac{r_1^2}{q_{2+}}-
\frac{q_1^2r_1^2}{q_{1+}q_{2+}r_{1-}}\Big)+
n^-_\mu\Big(\frac{q_2^2r_{1-}}{r_{2-}p_-}-2\frac{q_1^2}{r_{1-}}
-\frac{q_2^2r_1^2}{q_{1+}r_{2-}p_-}\Big).
\eeq
On mass shell, multiplied by the polarization vector,
\beq
X_{3\ep}=2(p\ep)_\perp\Big(\frac{q_1^2}{p_+r_{1-}}
+\frac{q_2^2r_{1-}}{r_{2-}p_\perp^2}-
\frac{q_2^2r_1^2}{q_{1+}r_{2-}p_\perp^2}\Big).
\eeq

{\bf 4.} ${\cal A}_4$

Again we present
\beq
{\cal A}_4=g^3C_4\frac{1}{t_1^2}X_4.
\eeq
We find
\beq
X_{4\mu}=-n^+_\mu\Big(2\frac{r_1^2}{q_{1+}}+\frac{r_1^2q_{1+}}{p_+q_{2+}}-
\frac{q_1^2r_1^2}{p_+q_{2+}r_{1-}}\Big)+
n^-_\mu\Big(\frac{q_1^2}{r_{1-}}+\frac{q_2^2}{r_{2-}}
-\frac{q_2^2r_1^2}{q_{1+}r_{1-}r_{2-}}\Big).
\eeq
On mass shell, multiplied by the polarization vector,
\beq
X_{4\ep}=-\frac{(p\ep)_\perp}{p_+}\Big(\frac{q_1^2}{r_{1-}}
+\frac{q_2^2}{r_{2-}}-
\frac{q_2^2r_1^2}{q_{1+}r_{1-}r_{2-}}\Big).
\eeq

{\bf 5.} ${\cal A}_5$

Presenting
\beq
{\cal A}_5=g^3C_5\frac{1}{t_1^2t_2^2}X_5,
\eeq
we use Eq. (\ref{a50}). To simplify quite cumbersome expressions we
from the start restrict ourselves
to mass-shell and multiply ${\cal A}_5$ by the polarization vector.
Singular terms are contained in $A^{(i)}$,
$i=1,2,3,5$ (term with $A^{(4)}$ drops out in our gauge).
We find
\[
A^{(1)}=2\frac{q_2^2r_{1-}}{r_{2-}}-2\frac{r_2^2q_{1+}}{q_{2+}},\ \
A^{(2)}=-2\frac{q_1^2r_{2-}}{r_{1-}}+2\frac{r_1^2q_{2+}}{q_{1+}},\]
\[
A^{(3)}=-\frac{q_1^2r_{2-}}{r_{1-}}-\frac{r_1^2q_{2+}}{q_{1+}}
-\frac{q_2^2r_{1-}}{r_{2-}}-\frac{r_2^2q_{1+}}{q_{2+}}+
\frac{q_1^2r_2^2}{q_{2+}r_{1-}}+\frac{q_2^2r_1^2}{q_{1+}r_{2-}}\]
and the most complicated term
\[A^{(5)}=
\frac{q_2^2}{r_{2-}}(a^{(1)},p+t_2)+\frac{q_1^2q_2^2}{r_{2-}}
+\frac{q_2^2r_1^2}{\rb}\]\[
+4\frac{q_1^2\qa\rb}{\ra}+4\frac{r_2^2\qa^2}{\qb}+
4\frac{q_1^2\qb\rb}{\ra}
+2\frac{q_2^2r_1^2\qb}{\qa\rb}-(1\lra 2),
\]
where $a^{(1)}=q_1+r_1$ and $a^{(2)}=q_2+r_2$
The coefficients are
\[(a^{(1)}\ep)=(a^{(1)},\ep)_\perp-(p\ep)_\perp\frac{q_{1+}}{p_+},\ \
(a^{(2)}\ep)=(a^{(2)},\ep)_\perp-(p\ep)_\perp\frac{q_{2+}}{p_+},\]
\[(\tau^{(3)}\ep)=(q_1-r_1-q_2+r_2,\ep)_\perp-
(p\ep)_\perp\frac{q_{1+}-q_{2+}}{p_+}.\]

Collecting all terms we find
\[
X_{5\ep}=(a^{(1)}\ep)_\perp A^{(1)}+(a^{(2)}\ep)_\perp A^{(2)}
+(t_1-t_2,\ep)_\perp A^{(3)}-\frac{(p\ep)_\perp}{p_+} A ,
\]
where
\[A=\frac{q_2^2}{\rb}(a^{(1)},p+t_2)+\frac{q_1^2q_2^2}{r_{2-}}+
2\frac{q_1^2q_2^2}{r_{2-}}\]\[
+3\frac{q_1^2\qa\rb}{\ra}+
\frac{r_2^2\qa^2}{\qb}+3\frac{q_1^2\qb\rb}{\ra}+
\frac{q_2^2r_1^2\qb}{\qa\rb} -(1\lra 2).\]

As we see each of the amplitudes ${\cal A}_i$, $i=1,...,5$ contains
both single poles in longitudinal momenta and double poles in
longitudinal momenta of in-coming and out-going reggeons.
Can they cancel in their sum together with terms obtained by
permutation of reggeons? The answer is negative, since different
amplitudes contain different denominators which moreover change with
permutation of reggeons. These denominators depend on transverse momenta
and so have different values. Therefore, at least at fixed transverse
momenta the pole singularities contain different
(and varying) coefficients in  amplitudes ${\cal A}_i$, $i=1,...,5$ and
the amplitudes obtained from them by permutations of reggeons.
So unlike the case of single projectile, in  production amplitudes
with two projectiles and targets
poles in the longitudinal momenta remain uncanceled, which
requires formulation of the way to do the longitudinal
integrations. Integration in the principal value sense is an obvious
choice.

\section{Conclusions}
We have derived the expression for the vertex RR$\to$RRP describing
gluon production in interaction of two in-coming and two out-going
reggeons. The vertex can be used for calculations of inclusive
cross-sections for gluon jet production in collision of a pair of
projectile nucleons with a pair of target nucleons and also of the
diffractive gluon jet production in deuteron-proton collisions.
The vertex turns out to be quite complicated but amenable to further
analytic and numerical calculations, which we postpone for future
publications.

A few important properties of the obtained vertex have been
demonstrated. The vertex is transversal in accordance with the gauge
invariance. It vanishes when one of the longitudinal momentum goes to
infinity, which allows to subsequently do  integrations over
longitudinal momenta in applications.

The vertex contains pole singularities at zero values of longitudinal
momenta inherited from intermediate induced vertices in the framework
of effective action. In the spirit of this framework one should consider
these poles in the principal value sense. Note that in contrast to
gluon production on several centers by a single projectile, where
rescattering effects cancel these poles, in amplitudes containing the
vertex RR$\to$RRP, like shown in Fig. \ref{fig1}, there are no
additional rescattering contributions, so that the mentioned
pole singularities are preserved in the amplitudes and should be taken
into account in longitudinal integrations.

Finally, again in contrast
to the case of a single projectile ~\cite{ref1,ref2,ref3}, we find that
the structure of the on-mass-shell vertex remains quite complicated and
cannot be restored
from the purely transverse picture, which is obtained by taking multiple
cuts of the amplitude ~\cite{bartels}. We believe that this is due to the
fact that the amplitude possesses additional singularities, apart from
the standard ones corresponding to physical intermediate gluons.

\section{Acknowledgements}
The authors acknowledge Saint-Petersburg State University
for a research grant 11.38.223.2015.
This work has been also supported by the RFFI grant 15-02-02097-a.

\section{Appendix. An alternative form of the amplitude ${\cal A}_5$}
Here we present a more explicit form for the amplitude ${\cal A}_5$
in terms of coefficients in the two Lipatov vertices $a_1,b_1,c_1$
and $a_2,b_2,c_2$.

We rewrite the triple gluon vertex as
\[
\Gamma_{\nu_1\mu,\nu_2}(t_1,p,t_2)=
gf^{d_1c,d_2}\Big(\ta_{\nu_2}g_{\mu\nu_1}+
\tb_{\nu_1}g_{\mu\nu_2}+\tc_\mu g_{\nu_1\nu_2}\Big),
\]
where
$\ta=-t_1-p,\ \ \tb=p+t_2,\ \ \tc=t_1-t_2$.
This allows to write the final vertex as
\beq
{\cal A}_{5\mu}=g^3C_5
\frac{1}{4(t_1^2+i0)(t_2^2+i0)}
\Big(a_{1\mu} A^{(1)}+a_{2\mu}  A^{(2)}+\tc_\mu  A^{(3)} + n^+_\mu A^{(4)}+
n^-_\mu A^{(5)}\Big),
\label{a50}
\eeq
where
\[ A^{(1)}=(a_2\ta)+b_2\ta_++c_2\ta_-\]
\[ A^{(2)}=(a_1\tb)+b_1\tb_++c_1\tb_-\]
\[ A^{(3)}=(a_2a_1)+b_1a_{2+}+c_1a_{2-}++b_2a_{1+}+c_2a_{1-}+b_2c_1+b_1c_2\]
\[ A^{(4)}=b_1(a_2\ta)+b_2(a_1\tb)+b_1b_2(\ta+\tb)_++c_2b_1\ta_-+b_2c_1\tb_-\]
\[ A^{(5)}=c_1(a_2\ta)+c_2(a_1\tb)+c_1c_2(\ta+\tb)_-+c_2b_1\tb_++
b_2c_1\ta_+\ .\]


\begin{thebibliography}{100}
%
\bibitem{BFKL1} V.S.Fadin, E.A.Kuraev, L.N.Lipatov, Phys. Lett
{\bf B 60}(1975) 50.
%
\bibitem{BFKL2} I.I.Balitsky, L.N.Lipatov, Sov. J. Nucl. Phys.
{\bf 28} (1978) 822.
%
\bibitem{bal} I.I.Balitsky, Nucl. Phys. {\bf B 463} (1996) 99.
%
\bibitem{kov} Yu.V.Kovchegov, Phys. Rev. {\bf D 60} (1999) 034008.
%
\bibitem{bra1} M.A.Braun, Phys. Lett. {\bf B 632} (2006) 297.
%
\bibitem{gelis1} F.Gelis, T.Lappi, R.Venugopalan, Phys. Rev. {\bf D 78}
(2008) 054019.
%
\bibitem{gelis2} F.Gelis, T.Lappi, R.Venugopalan, Phys. Rev. {\bf D 78}
(2008) 054020.
%
\bibitem{gelis3} F.Gelis, T.Lappi, R.Venugopalan, Phys. Rev. {\bf D 79}
(2009) 094017.
%
\bibitem{dusling} K.Dusling, F.Gelis, T.Lappi, R.Venugopalan,
Nucl. Phys. {\bf A 836} (2010) 159.
%
\bibitem{kov1} Yu.V.Kovchegov, Nucl. Phys. {\bf A 692} (2001) 567.
%
\bibitem{bal1} I.I.Balitsky, Phys. Rev. {\bf D 72} (2005) 074027.
%
\bibitem{dd} M.A.Braun, Eur. Phys. J. {\bf C 73} (2013) :2418.
%
\bibitem{odderon} J.Bartels, V.S.Fadin, L.N.Lipatov, G.P.Vacca,
Nucl.Phys. {\bf B 867 } (2013) 827;\\ arXiv:1210.0797.
%
\bibitem{lip} L.N.Lipatov, Nucl. Phys. {\bf B 452} (1995) 369.
%
\bibitem{bravyaz} M.A.Braun, M.I.Vyazovsky, Eur. Phys. J.
{\bf C 51} (2007) 103.
%
\bibitem{ref1} M.A.Braun, L.N.Lipatov, M.Yu.Salykin, M.I.Vyazovsky,
\\ Eur. Phys. J. {\bf C 71} (2011) :1639.
%
\bibitem{ref2} M.A.Braun, S.S.Pozdnyakov, M.Yu.Salykin, M.I.Vyazovsky,
\\ Eur. Phys. J. {\bf C 72} (2012) :2223.
%
\bibitem{ref3} M.A.Braun, S.S.Pozdnyakov, M.Yu.Salykin, M.I.Vyazovsky,
\\ Eur. Phys. J. {\bf C 74} (2014) :2989.
%
\bibitem{bartels} J. Bartels, Nucl. Phys. {\bf B 175} (1980) 365.
\end{thebibliography}
\end{document}